\documentclass[prl,superscriptaddress,twocolumn,amsmath,showpacs]{revtex4-1}

\usepackage{bm}
\usepackage{graphicx}
\usepackage[usenames,dvipsnames,svgnames,table]{xcolor}
\usepackage[colorlinks=true,linkcolor=RoyalBlue,citecolor=RoyalBlue]{hyperref}
\usepackage{braket}
\usepackage{mathrsfs}\usepackage{breakurl}\usepackage{feynmp}\usepackage{subfigure}\usepackage{color}

\begin{document}

\title{RKKY interaction of magnetic impurities in Dirac and Weyl semimetals}  

\author{Hao-Ran Chang}
\affiliation{Department of Physics and Institute of Solid State Physics, Sichuan Normal University, Chengdu, Sichuan 610066, China}
\author{Jianhui Zhou}
\email{jianhuizhou1@gmail.com}
\affiliation{Department of Physics, Carnegie Mellon University, Pittsburgh, Pennsylvania 15213, USA}
\author{Shi-Xiong Wang}
\affiliation{Department of Physics and Institute of Solid State Physics, Sichuan Normal University, Chengdu, Sichuan 610066, China}
\author{Wen-Yu Shan}
\affiliation{Department of Physics, Carnegie Mellon University, Pittsburgh, Pennsylvania 15213, USA}

\author{Di Xiao}
\email{dixiao@cmu.edu}
\affiliation{Department of Physics, Carnegie Mellon University, Pittsburgh, Pennsylvania 15213, USA}

\date{\today}

\begin{abstract}
We theoretically study the Ruderman-Kittel-Kasuya-Yosida (RKKY) interaction between magnetic impurities in both Dirac and Weyl semimetals (SMs).  We find that the internode process, as well as the unique three-dimensional spin-momentum locking, has significant influences on the RKKY interaction, resulting in both a Heisenberg and an Ising term, and an additional Dzyaloshinsky-Moriya term if the inversion symmetry is absent.  These interactions can lead to rich spin textures and possible ferromagnetism in Dirac and time-reversal symmetry-invariant Weyl SMs.
The effect of anisotropic Dirac and Weyl nodes on the RKKY interaction is also discussed.
Our results provide an alternative scheme to engineer topological SMs and shed new light on the application of Dirac and Weyl SMs in spintronics.
\end{abstract}


\maketitle

%
Three-dimensional (3D) Dirac semimetals (SMs)~\cite{SMYoung} are topological states of matter and can be seen as bulk analog of graphene. Their conduction and valence bands with linear dispersion
touch each other at a finite number of points, called the Dirac nodes, in the 3D Brillouin zone.
Dirac nodes are fourfold degenerate, protected by both time-reversal symmetry (TRS) and inversion symmetry.
Breaking either symmetry in Dirac SMs leads to Weyl SMs~\cite{XGWanPRB}, which host Weyl nodes.
These Weyl nodes can be viewed as effective magnetic monopoles in the momentum space~\cite{Volovik}, acting as the source
and drain of the Berry curvature field~\cite{XiaoRMP}. This nontrivial topology can lead to exotic superfluid~\cite{excitonSF} and superconducting phases~\cite{SC}, unique Fermi arc state~\cite{SYXUFermiArc}, helical spin order~\cite{XQSun}, various novel electromagnetic responses,
such as the chiral anomaly~\cite{HosurCRP,Hughes}, the chiral magnetic effect~\cite{CME}, negative magnetoresistance~\cite{NMR}, and the chiral Hall effect~\cite{ChiralHall}.

Many aforementioned exotic phenomena rely on the separation of Weyl nodes in momentum space due to the intrinsic TRS breaking.
So far, angle-resolved photoemission spectroscopy and magnetotransport measurements have identified
(Bi$_{1-x}$In$_x$)$_{2}$Se$_3$~\cite{BrahlekBiInSe},
Na$_{3}$Bi~\cite{ZKLiuNaBi}, and Cd$_3$As$_{2}$~\cite{BorisenkoCdAs} as Dirac SMs,
and noncentrosymmetric transition-metal monosphides TaAs, NbAs, NbP and TaP as Weyl SMs~\cite{WengTaAs,TaAsXu,TaAsLv,NbAs, NbP, TaPXu}.
There are, however, few experimental realizations of TRS breaking Weyl SMs without Landau quantization~\cite{YbMnBi}.
Even though a magnetic field can break the TRS,
it inevitably couples to both the spin and orbital motion of electrons through Zeeman splitting and Landau quantization, respectively.
On the other hand, magnetic doping technique has recently been utilized in experimental implementation of the quantum anomalous Hall effect in thin films of topological insulators~\cite{QAH}.
Thus, one may naturally wonder whether Weyl SMs without TRS can emerge from  Dirac and time-reversal invariant Weyl SMs
through introducing magnetic dopants rather than applying a magnetic field.

Motivated by the above observation, in this Rapid Communication we study the Ruderman-Kittel-Kasuya-Yosida (RKKY) interaction between magnetic dopants in both Dirac and Weyl SMs.  We find that the internode process, as well as the unique 3D spin-momentum locking, has significant influences on the RKKY interaction, resulting in both a Heisenberg and an Ising term, and an additional Dzyaloshinsky-Moriya term if the inversion symmetry is absent.  These interactions can lead to rich spin textures and possibly ferromagnetism in Dirac and TRS-invariant Weyl SMs.
The effect of anisotropic Dirac and Weyl nodes on the RKKY interaction is also discussed.
Our results provide an alternative scheme to engineer topological SMs and shed new light on the application of Dirac and Weyl SMs in spintronics.
%

%
In general, a pair of Weyl nodes of opposite chirality can be described by the following Hamiltonian
\begin{equation}
H_{0}=\chi \big[v_{F}\boldsymbol{\sigma} \cdot \left({\bm k} - \chi \bm{Q} \right)
+\sigma_{0} {Q}_{0}\big] \;,
\label{HDW}
\end{equation}
where $\bm k$ is the wave vector, $v_{F}$ is the Fermi velocity, and $\chi = \pm1$ refers to the chirality of the Weyl nodes.
If $(\bm Q,Q_0)=0$, the two Weyl nodes overlap with each other and the Hamiltonian $H_0$ describes Dirac SMs~[see Fig.~1(a)].
For noncentrosymmetric Weyl SMs that preserve the TRS, $\bm Q$ must be zero and $Q_0$ can be nonzero.  As a result, the two Weyl nodes are located at the same $\bm k$ point but can have different energies as shown in Fig.~1(b).
Hence for a given carrier density, there are two unequal Fermi wave vectors.
On the other hand, for Weyl SMs with broken TRS but with inversion symmetry, we have $Q_0 = 0$ and $\bm Q \neq 0$; the two Weyl modes have the same energy but
reside at different $\bm k$ points $\pm\bm{Q}$ in the Brillouin zone~[see Fig.~1(c)].
Here the Pauli matrices $\bm \sigma = (\sigma_x, \sigma_y, \sigma_z)$
refer to the real spin degree of freedom of electrons.  They may also refer to pseudospin degree of freedom, which will be discussed later.
%
\begin{figure}
\includegraphics[scale=0.35]{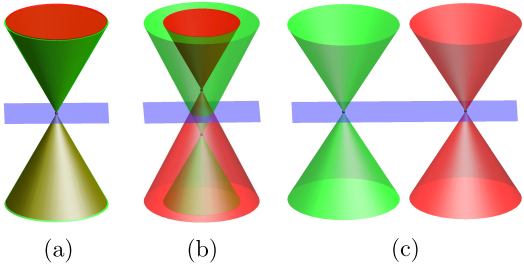}
\caption{(Color online)
Low energy spectra for Dirac SMs~(a), noncentrosymmetric Weyl SMs~(b), and Weyl SMs
without TRS~(c), respectively.
The color of cone indicates the chirality of Weyl nodes, while the blue plane is the Fermi level $\varepsilon_{F}=0$.}
\label{figdisp}
\end{figure}

%
We assume that the interaction between the 3D itinerant
Weyl fermions and a magnetic impurity ${\bm S}_{i}$ located at $\bm R_i$ can be expressed as the standard $s$-$d$ interaction Hamiltonian
$H_{I}=(J \tau_0+\lambda  \tau_x){\bm S}_{i}\cdot \bm \sigma \delta({\bm r} - {\bm R}_{i})$,
where $J$ and $\lambda$ refer to the strength of the $s$-$d$ exchange interaction in
the intranode process and the internode process, respectively.
The identity matrix $\tau_0$
and Pauli matrix
$\tau_x$ act on the chirality space.

According to the second order perturbation theory~\cite{Mahan}, at zero temperature the RKKY interaction
between two magnetic impurities mediated by Weyl fermions is given by
$H_{\text{RKKY}} = - \frac{1}{\pi}\mathrm{Im} \int_{-\infty}^{\varepsilon_{F}}d\varepsilon\,
\mathrm{Tr}[(J \tau_{0}+\lambda \tau_{x}) \bm{S}_1 \cdot \bm \sigma \mathcal{G} ({\bm R} ;\varepsilon) (J \tau_{0}+\lambda \tau_{x}) {\bm S}_2 \cdot \bm \sigma \mathcal{G} (-\bm R ; \varepsilon)]$ with ${\bm R}={\bm R}_{2}-{\bm R}_{1}$,
where  $\varepsilon_{F}$ is the Fermi energy and Tr means a trace over the spin and pseudospin degree of freedom of itinerant Weyl fermion.  The Green's function in the energy-coordinate representation is given as
$\mathcal{G} ({\bm R} ; \varepsilon) = G_{+} ({\bm R} ; \varepsilon) \oplus G_{-} (\bm R ; \varepsilon)$.  After some algebra, we find the RKKY interaction~\cite{footnote}
\begin{alignat}{1}
&H_{\text{RKKY}}=\sum_{\alpha,\beta,\chi,\chi^{\prime}}
\left[J^{2}\delta_{\chi\chi^{\prime}}+\lambda^{2}
(1-\delta_{\chi\chi^{\prime}})\right]S_{1}^{\alpha}S_{2}^{\beta}\nonumber\\
&\times\mathrm{Im}\left\{-\frac{1}{\pi}\int_{-\infty}^{\varepsilon_{F}}d\varepsilon
\mathrm{Tr}[\sigma_{\alpha}G_{\chi} (\bm{R};\varepsilon)
\sigma_{\beta}G_{\chi^{\prime}} (-\bm{R};\varepsilon)]\right\} \;,
\label{RKKY}
\end{alignat}
which includes contributions from both the intranode process and the internode process.
%
%

We now apply the above formalism to study the RKKY interaction in Dirac and Weyl SMs.  We note that close to the Weyl nodes, the effective Hamiltonian is proportional to $\bm k \cdot \bm {\sigma}$,
which can be viewed as the 3D counterpart of graphene in real spin space.
It is in contrast with the surface state of 3D topological insulators (TIs)~\cite{QinLiuTI,JJZhuTI,Biswas},
in which spin and velocity are perpendicular to each other.  As we show below, this kind of hedgehog spin texture around Weyl nodes (spin is aligned with momentum) can have significant effects  on the magnetism of magnetic impurities.

Let us first consider the isotropic Dirac SMs, described by
$H_{\mathrm{D}}({\bm k}) = \chi v_F \bm k \cdot \bm \sigma$ with the energy dispersion $\varepsilon_{k}=\pm v_{F} k$.  The effect of anisotropic energy dispersion will be discussed later.
The Green's function corresponding to $H_{\mathrm{D}}({\bm k})$ in momentum space takes the form
$G_{\chi}^{-1}({\bm k};\varepsilon)=(\varepsilon+i\eta)\sigma_{0}-H_{D}({\bm k})$, where $\eta$ is a positive infinitesimal.
$G_{\chi}(\pm{\bm R};\varepsilon)$ can be obtained from $G_{\chi}({\bm k};\varepsilon)$
through integrating over the momentum near the Weyl node $\chi$,
$G_{\chi}(\pm{\bm R};\varepsilon)
=\int\frac{d^{3}{\bm k}}{(2\pi)^{3}}G_{\chi}({\bm k};\varepsilon)
\exp\left(\pm i{\bm k}\cdot{\bm R}\right)$.
Carrying out the integration over $\bm k$ leads to (more details are presented in the Supplemental Material~\cite{SM})
$G_{\chi}(\pm{\bm R};\varepsilon)=\sigma_{0} G_{0}(R;\varepsilon) \pm \chi\sigma_{j} G_{R}(R;\varepsilon)$,
where the Green's functions are defined as $G_{0}(R;\varepsilon)
=\frac{-\varepsilon}{4\pi v_{F}^{2}R}\exp[i\xi_{\varepsilon}]$ and
$G_{R}(R;\varepsilon)=\frac{-i}{4\pi v_{F}R^{2}}(1-i\xi_{\varepsilon}) \exp[i\xi_{\varepsilon}]$,
with the dimensionless parameter $\xi_{\varepsilon}=\varepsilon R/v_{F}$.
In the above calculation, we take ${\bm R}$ to be aligned to the $j$ axis, i.e., ${\bm R}=R\boldsymbol{e}_{j}$.

After lengthy but straightforward calculations, we obtain the RKKY interaction for the Dirac SMs
(the detailed calculation can be found in the Supplemental Material~\cite{SM})
\begin{alignat}{1}
H_{\mathrm{RKKY}}^{\mathrm{D}} & =F_{\mathrm{H}}^{\mathrm{D}}(R,\xi_{F})\boldsymbol{S}_{1}\cdot\boldsymbol{S}_{2}+F_{\mathrm{Ising}}^{\mathrm{D}}(R,\xi_{F})S_{1}^{j}S_{2}^{j} \;,
\end{alignat}
where the range functions for the Heisenberg and Ising terms are given by
\begin{align}\nonumber
F_{\mathrm{H}}^{\mathrm{D}}(R,\xi_{F}) & =-\left\{ J^{2}[(3-2\xi_{F}^{2})\cos(2\xi_{F})+4\xi_{F}\sin(2\xi_{F})]\right.\\
 & \left.-2\lambda^{2}[\cos(2\xi_{F})+\xi_{F}\sin(2\xi_{F})]\right\} /8\pi^{3}v_{F}R^{5} \label{FH}\\\nonumber
F_{\mathrm{Ising}}^{\mathrm{D}}(R,\xi_{F}) & =-(J^{2}-\lambda^{2})\left[(2\xi_{F}^{2}-5)\cos(2\xi_{F})\right.\\  \label{FI}
 & \left.-6\xi_{F}\sin(2\xi_{F})\right]/8\pi^{3}v_{F}R^{5} \;,
 \end{align}
with $\xi_{F}=\varepsilon_{F}R/v_{F}$. Here the superscript $\mathrm{D}$ stands for Dirac SMs.
One can clearly see that if $J^2 = \lambda^2$, the contribution to the Ising term from the internode process
cancels that from the intranode process exactly, i.e., $F_{\mathrm{Ising}}^{\mathrm{D}}=0$.
This cancellation is due to the restoration of the spin rotation symmetry by the internode process.

At long range ($\xi_{F} \gg1$) and for finite $\varepsilon_F$, the RKKY interaction reduces to a simple form
\begin{alignat}{1}
H_{\mathrm{RKKY}}^{\mathrm{D}} & \approx\frac{J^2 \varepsilon_{F}^{2} \cos(2\xi_{F})}{4\pi^{3}v_{F}^{3}R^{3}} \boldsymbol{S}_{1}\cdot\boldsymbol{S}_{2} \;,
\label{DLong}
\end{alignat}
which reflects the long-range and oscillatory nature of the RKKY interaction.
The ferromagnetism (FM) or antiferromagnetism (AFM) of magnetic impurities depends
on both the concentration of impurities and the carrier density of Dirac fermions through $R$ and $\varepsilon_{F}$, respectively.

Interestingly, for the intrinsic case ($\varepsilon_{F} = 0$), the RKKY interaction becomes nonoscillatory:
\begin{alignat}{1}
H_{\mathrm{RKKY}}^{\mathrm{D}} & =\frac{-(3J^{2}-2\lambda^2)}{8\pi^{3}v_{F}R^{5}} \boldsymbol{S}_{1}\cdot\boldsymbol{S}_{2} \;.
\label{DIntrinsic}
\end{alignat}
We can see that if $J^2 = \lambda^2$, the exchange coupling is always ferromagnetic.  For sufficiently large impurity density, this will lead to a spontaneous magnetization, which then drives Dirac SMs into Weyl SMs with broken TRS.
Consequently, a nonzero anomalous Hall conductivity proportional to the separation of the two Weyl nodes in momentum space will appear~\cite{AHEYang}.
The ferromagnetic transition temperature $T_c$ depends on the specifics of both magnetic impurities and host materials,
which need detailed first-principles studies~\cite{Jungwirth}.
It is clear that its spatial dependence as $1/R^5$ differs from $1/R^3$ for the intrinsic graphene \cite{Kogan}.
%

\begin{figure}
\includegraphics[scale=0.4]{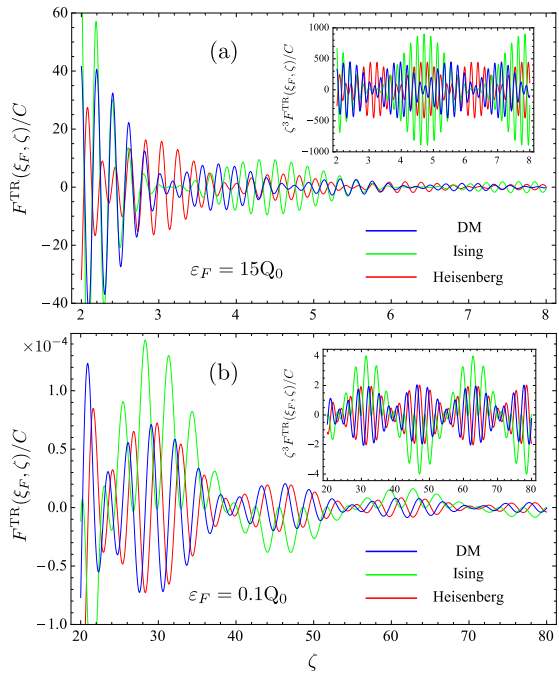}
\caption{(Color online)
The exact range functions of RKKY interaction of the noncentrosymmetric Weyl SMs ($F^{\mathrm{TR}}/C$) as a function of
the reduced parameter $\zeta$ at the Fermi energy $\varepsilon_{F}=15Q_{0}$ (a) and
$\varepsilon_{F}=0.1Q_{0}$ (b), with $C=-J^2Q_{0}^5/(2\pi v_{F}^2)^3$.
The inset in each panel for $\zeta^3 F^{\mathrm{TR}}/C$ shows an evident beating feature for each term in this RKKY interaction.}
\label{ncwrgf}
\end{figure}

Next we consider noncentrosymmetric Weyl SMs, in which the two Weyl nodes with opposite chirality
have different energies ($Q_{0}\neq0$). For a given carrier density, the magnitudes of Fermi wave vectors of
these two Weyl nodes are distinct, therefore $\xi_{F}^{+}\neq \xi_{F}^{-}$ with $\xi_{F}^{\chi}=(\varepsilon_{F}-\chi Q_{0})R/v_{F}$.
Following the same procedure, we obtain the RKKY interaction for the noncentrosymmetric Weyl SMs
\begin{alignat}{1} \label{RFTR}
H_{\mathrm{RKKY}}^{\mathrm{TR}} & =F_{\mathrm{H}}^{\mathrm{TR}}(\xi_{F},\zeta)\boldsymbol{S}_{1}\cdot\boldsymbol{S}_{2}+F_{\mathrm{Ising}}^{\mathrm{TR}}(\xi_{F},\zeta)\nonumber \\
 & \times S_{1}^{j}S_{2}^{j}+F_{\mathrm{DM}}^{\mathrm{TR}}(\xi_{F},\zeta)(\boldsymbol{S}_{1}\times\boldsymbol{S}_{2})_{j} \;,
\end{alignat}
with $\zeta=Q_0 R/v_{F}$ and
the $j$th component of the spin $S^j$ is along the direction connecting the two impurities. Here the superscript TR stands for the TRS invariant Weyl SMs.
We can see that the RKKY interaction consists of three terms, namely, a Heisenberg term, an Ising term, and a Dzyaloshinsky-Moriya (DM) term~ \cite{DMInt}.  This is
similar to the surface state of TIs~\cite{JJZhuTI}.
The specific expressions of these range functions $(F_{\mathrm{H}}^{\mathrm{TR}},F_{\mathrm{Ising}}^{\mathrm{TR}},F_{\mathrm{DM}}^{\mathrm{TR}})$
are given in the Supplemental Material \cite{SM}.

Figure~2 shows the range functions to two sets of parameters.
As we can see, these range functions display a damped oscillatory behavior
with increasing distance $R$, with each term dominating in different regimes
of the parameters $\varepsilon_{F}$ and $R$.
In addition, these range functions
oscillate with two distinct periods and form a beating pattern.
The beating feature, originated from the two unequal Fermi wave vectors,  manifests
itself by multiplying each of the range functions in Eq.~(\ref{RFTR}) by $\zeta^3$ as
shown in the insets of Figs.~2(a) and 2(b).
The measurement of beating period can be used to determine the energy
difference~$Q_0$~\cite{Beating}.  This beating structure does not occur in the typical surface state of TIs where there is only one Fermi circle~\cite{QinLiuTI,JJZhuTI,Biswas}.

The DM term is essential for realizing spiral spin states and skyrmions, and also has potential applications in spintronics.
It may also provide some hint to understand the recent experimental observation of the TRS-breaking Weyl metal in YbMnBi$_2$~\cite{YbMnBi}.
In this material, although there is a global inversion symmetry, in each layer the inversion symmetry is broken, which could give rise to a nonvanishing DM term.
The combination of the DM interaction and the AFM order of Mn can cause a canted AFM order with a nonvanishing net magnetization observed in the experiment~\cite{YbMnBi}.
%

Finally, we discuss TRS-breaking Weyl SMs, in which the Weyl nodes with opposite chirality reside at different $\bm k$ points in momentum space $\pm \bm{Q}$.
We further assume the inversion symmetry remains intact, thus the DM term does not appear.
We take into account the effect of the separation of Weyl nodes in the internode process.
The corresponding RKKY interaction can be obtained from that of Dirac SMs in Eqs.~(\ref{FH}) and (\ref{FI}) by replacing
$\lambda^2$ with $\lambda^2 \cos(2 \bm{Q} \cdot \bm R)$.
It can be seen that the internode process gives rise to an oscillating term proportional to $\sim\cos(2{\bm Q}\cdot{\bm R})$, which is absent in the typical surface state of TIs~\cite{QinLiuTI,JJZhuTI}.
Because of the long-range and oscillatory nature of the RKKY interaction,
for a large momentum separation $2 \bm{Q}$, the part from the internode process will vanish after averaging over position and thus does not contribute to the net magnetization.
Therefore, the intranode process dominates the RKKY interaction in this case.
At long distance, we have the approximate RKKY interaction
\begin{alignat}{1}
H_{\mathrm{RKKY}}^{\mathrm{I}} \approx \frac{J^2 \varepsilon_{F}^{2} \cos(2\xi_{F})}{4\pi^{3}v_{F}^{3}R^{3}}
(\boldsymbol{S}_{1}\cdot\boldsymbol{S}_{2}-S_{1}^{j}S_{2}^{j}) \;,
\end{alignat}
where the superscript I refers to the centrosymmetric Weyl SMs.
Since the asymptotic range functions for the Heisenberg and Ising terms exhibit the same oscillatory
behavior but with a phase differing by $\pi$, the Ising term always cancels the $j$th component
of the Heisenberg term, leading to the $XY$-like spin model for $j=z$.
On the other hand, for the intrinsic case the RKKY interaction becomes
\begin{alignat}{1}
H_{\mathrm{RKKY}}^{\mathrm{I}}=\frac{-J^2}{8\pi^{3}v_{F}R^{5}}
(3 \boldsymbol{S}_{1}\cdot\boldsymbol{S}_{2}-5 S_{1}^{j}S_{2}^{j}) \;,
\end{alignat}
which could realize various spin models, such as the $XXZ$-like spin model for $j=z$.
The resulting spin configurations of impurities ($XY$- and $XXZ$-like spin models) can be accessed by a variety of experimental techniques,
such as neutron scattering technique.

In reality, almost all experimental realizations of Weyl or Dirac SMs
(and theoretically conjectured systems, such as~\cite{AHEYang,GXu,ZJW1,ZJW2})
possess a strongly anisotropic single-particle dispersion.
Here we consider the effect of anisotropy on the RKKY interaction.
The effective Hamiltonian for fermions near the anisotropic Weyl node $\chi$ takes the form
$\tilde{H}_{0}=\chi(v_{x}\tilde{k}_{x}\sigma_{x}+v_{y}\tilde{k}_{y}\sigma_{y}+ v_{z}\tilde{k}_{z}\sigma_{z}$),
where the Fermi velocities $v_{\alpha}$ in different direction are different, $ |v_{x}| \neq |v_{y}| \neq |v_{z}|$.
To simplify our discussion, we set $v_j >0$ with $j = x,y,z$.
The corresponding energy spectrum is given as
$\varepsilon_{\tilde{k}}=\pm(v_{x}^{2}\tilde{k}_{x}^{2}+v_{y}^{2}\tilde{k}_{y}^{2}
+v_{z}^{2}\tilde{k}_{z}^{2})^{1/2}$.
It is instructive to make the following transformation as $(v_{i}\tilde{Q}_{i},v_{i}\tilde{k}_{i},v_{F}\tilde{R}_{i})
\equiv(v_{F}Q_{i},v_{F}k_{i},v_{i}R_{i})$, such that we
relate the Green's function of an anisotropic Weyl node in the energy-coordinate representation to that of an isotropic one as
$\tilde{G}_{\chi}(\pm \tilde{{\bm R}};\varepsilon)
=\lambda_{A} G_{\chi}(\pm {\bm R};\varepsilon)$,
with $\lambda_{A}=v_{F}^{3}/(v_{x}v_{y}v_{z})$.
We note that the above transformation converts a Fermi elliptic sphere into a Fermi sphere but preserves the volume of Fermi sphere.
This allows us to connect the RKKY interaction of anisotropic Dirac SMs to the counterpart of the isotropic ones
\begin{eqnarray}
\tilde{H}_{\mathrm{RKKY}}(\tilde{R})
=\lambda_{A}^{2} H_{\mathrm{RKKY}}(v_F \tilde R /v_{i}) \;,
\label{ARKKY}
\end{eqnarray}
which implies that the anisotropy of Dirac or Weyl nodes must lead to an anisotropic RKKY interaction.
It is clear that for $v_{x}=v_{y}=v_{z}$ ($\lambda_{A}=1$), the above expression in Eq.~(\ref{ARKKY}) can reduce to that of the isotropic one.
%

Before drawing conclusions, we briefly discuss the pseudospin case, in which
the Pauli matrices in Eq.~(\ref{HDW}) may refer to pseudospin degree of freedom such as orbital index. This pseudospin case is similar to graphene.
In the absence of the spin-momentum locking, the RKKY interaction only contains the conventional Heisenberg term~\cite{RKKY3D}.
Hence, the RKKY interaction directly associates with the Fourier transform of the static density-density response function~\cite{JHZPlasmon}.
We leave all the specific expressions of RKKY interaction to the Supplemental Material~\cite{SM}.
It should be emphasized that the RKKY interaction in this intrinsic Dirac SMs is also a nonoscillatory
Heisenberg term $H_{\mathrm{RKKY}}=\frac{(2 J^2-3 \lambda^2)}{4 \pi^3 v_F R^5}\boldsymbol{S}_{1} \cdot \boldsymbol{S}_{2}$,
which allows the spontaneous magnetization of magnetic impurities for $|\lambda| > (2/3)^{1/2} |J|$.
Compared with Eq.~(\ref{DIntrinsic}), the factor of $2$ comes from the degeneracy of real spin degree of freedom.
%

%
%
In summary, we have studied the RKKY interaction between magnetic impurities in Dirac and Weyl SMs.  We found that it is possible to realize a spontaneous magnetization in these systems.  The RKKY interaction in general contains the Heisenberg, Ising, and DM terms, which can give rise to rich spin textures of the impurities.
These findings provide an alternative scheme to engineer topological SMs and pave the way for the application of Dirac and Weyl SMs for spintronics.

\textit{Note~added}. Recently, a related paper appeared, in which some results on the anisotropic Weyl SMs have also been obtained \cite{Vahid}.
Also recently, the Kondo effect of a single magnetic impurity in the Weyl SMs has also been discussed in \cite{YiZhou,Mitchell}.

%
We are grateful to Ran Cheng and Shengyuan Yang for many valuable discussions and comments.
This work is supported by AFOSR (Grant No. FA9550-14-1-0277), NSF (Grant No.~EFRI-1433496), Natural Science Foundation of China (Grant No.~11547200), Natural Science Foundation of
Sichuan Educational Committee (Grant No.~13ZB0157), and Sichuan Normal University (Grant No.~15YB001).


\onecolumngrid

\section*{Supplementary material for"RKKY interaction of magnetic impurities in Dirac and Weyl semimetals"}

This supplemental material is organized as follows. In Sec. I, we present the general expression of RKKY interaction in terms of the Green function.  
In Sec. II, the Green function in the energy-coordinate representation is calculated detailedly.  
In Sec. III, we further calculate the RKKY interactions for 3D Dirac SMs,
Weyl SMs without time reversal symmetry (TRS), and noncentrosymmetric Weyl SMs in both cases with  and without spin-momentum locking.
In Sec. IV, we investigate the impact of anisotropy of Weyl or Dirac nodes on RKKY interaction. 
In Sec. V, we give some useful integrals for calculating the range functions.  
%

\section{Formalism of RKKY interaction}
 %
Following the second order perturbation theory~\cite{Mahan}, we express the general RKKY interaction
between two magnetic impurities mediated by itinerant fermions in terms of the Green function as
\begin{equation}
H_{\mathrm{RKKY}}=-\frac{1}{\pi}\mathrm{Im}\int_{-\infty}^{\varepsilon_{F}}d\varepsilon\,\mathrm{Tr}\Bigl[\left(J\tau_{0}+\lambda \tau_{x}\right) \boldsymbol{S}_{1}\cdot\boldsymbol{\sigma}\,\mathcal{G}\left(\mathbf{R},\varepsilon\right)\left(J\tau_{0}+\lambda \tau_{x}\right)\boldsymbol{S}_{2}\cdot\boldsymbol{\sigma}\,\mathcal{G}\left(-\mathbf{R},\varepsilon\right)\Bigr],\label{eq:rkkydef0}
\end{equation}
where $\varepsilon_{F}$ is the Fermi energy, $\mathrm{Tr}$ means a trace over the (pseudo)spin degree of freedom of itinerant Weyl fermion, the identity matrix $\tau_0$
and Pauli matrix $\tau_x$ act upon the chirality space, $J$ and $\lambda$ refer to
the strength of the s-d exchange interaction in the intranode process and the internode process, respectively.
The Green function in energy-coordinate representation is given by
\begin{equation}
\mathcal{G\left(\text{\ensuremath{\pm\mathbf{R}},\ensuremath{\varepsilon}}\right)}=\begin{pmatrix}G_{+}\mbox{(\ensuremath{\pm\text{\ensuremath{\mathbf{R}},\ensuremath{\varepsilon}}})} & 0\\
0 & G_{-}\mbox{(\ensuremath{\pm\text{\ensuremath{\mathbf{R}},\ensuremath{\varepsilon}}})}
\end{pmatrix},\label{eq:GreenR}
\end{equation}
where $G_{\chi}\left(\text{\ensuremath{\mathbf{\pm R}},\ensuremath{\varepsilon}}\right)$
is the Fourier transform of the Green function in momentum space $G_{\chi} (\mathbf{k},\varepsilon)$. Inserting Eq.~$(\ref{eq:GreenR})$ into Eq.~$(\ref{eq:rkkydef0})$,
we obtain the RKKY interaction
\begin{eqnarray}
 &  & H_{\mathrm{RKKY}}=\sum_{\chi,\chi^{\prime},\alpha,\beta}
 \Big[J^{2}\delta_{\chi\chi^{\prime}}+\lambda^{2}(1-\delta_{\chi\chi^{\prime}})\Big]S_{1}^{\alpha}S_{2}^{\beta}
 \mathrm{Im}\mathcal{F}_{\alpha\beta}^{\chi\chi^{\prime}}(R;\varepsilon_{F}),
 \label{eq:defrkky1}
\end{eqnarray}
where the functions in different channel are given by
\begin{eqnarray}
 &  & \mathcal{F}_{\alpha\beta}^{\chi\chi^{\prime}}(R;\varepsilon_{F})=-\frac{1}{\pi}\int_{-\infty}^{\varepsilon_{F}}d\varepsilon\mathrm{\, Tr}\left[\sigma_{\alpha}G_{\chi}\left(\text{\ensuremath{\mathbf{R}},\ensuremath{\varepsilon}}\right)\sigma_{\beta}G_{\chi^{\prime}}\left(-\text{\ensuremath{\mathbf{R}},\ensuremath{\varepsilon}}\right)\right].\label{eq:Fchialpha1}
\end{eqnarray}
It should be noted that Eq.~$\left(\ref{eq:defrkky1}\right)$ is our main expression of RKKY interaction, and the
evaluation of $\mathcal{F}_{\alpha\beta}^{\chi\chi^{\prime}}(R;\varepsilon_{F})$ is left to the next part.

\section{Derivation of the Green functions}  
\label{GFs}

We at first calculate the Green function of the Dirac SMs in energy-coordinate representation,
which is the Fourier transform of that in momentum space
$G_{\chi}^{-1}(\mathbf{k},\varepsilon)=\left(\varepsilon+i\eta\right)\sigma_{0}-\chi v_{F}\boldsymbol{\sigma}\cdot\mathbf{k}$,

\begin{align}
G_{\chi}\left(\text{\ensuremath{\mathbf{\pm R}},\ensuremath{\varepsilon}}\right) & =\int\frac{d^{3}\mathbf{k}}{\left(2\pi\right)^{3}}G_{\chi}\left(\mathbf{k},\varepsilon\right)\exp\bigl(\pm i\mathbf{k}\cdot\mathbf{R}\bigr)=\int\frac{d^{3}\mathbf{k}}{(2\pi)^{3}}\frac{\varepsilon\sigma_{0}\pm\chi v_{F}\boldsymbol{\sigma}\cdot\mathbf{k}}{(\varepsilon+i\eta)^{2}-\varepsilon_{k}^{2}}\exp\left(i\mathbf{k}\cdot\mathbf{R}\right),
\end{align}
where $\varepsilon_{k}=\pm v_{F}k$. In general,
the momentum $\mathbf{k}$ can be decomposed into the directions parallel and perpendicular
to $\boldsymbol{e}_{R}$ as $\mathbf{k}=\mathbf{k}_{\parallel}+\mathbf{k}_{\perp}\equiv(\mathbf{k}\cdot\boldsymbol{e}_{R})\boldsymbol{e}_{R}+(\boldsymbol{e}_{R}\times\mathbf{k})\times\boldsymbol{e}_{R}$
with $\boldsymbol{e}_{R}=\mathbf{R}/|\mathbf{R}|$. Because of $\mathbf{k}_{\perp}\cdot\boldsymbol{e}_{R}=0$,
the integration over angular variables causes the part of $\mathbf{k}_{\perp}$
in $G_{\chi}\left(\mathbf{R};\varepsilon\right)$ to vanish, we then have
\begin{equation}
G_{\chi}(\pm\mathbf{R},\varepsilon)\equiv\sigma_{0}G_{0}(R;\varepsilon)\pm\chi\boldsymbol{\sigma}\cdot\boldsymbol{e}_{R}G_{R}(R;\varepsilon).\label{eq:GchiR}
\end{equation}
Therefore we only need to evaluate two integrals, $G_{0}(R;\varepsilon)$ and $G_{R}(R;\varepsilon)$. The former is calculated as follows,
\begin{equation}
G_{0}(R;\varepsilon)=\varepsilon\int\frac{d^{3}\mathbf{k}}{(2\pi)^{3}}\frac{\exp(ikR\cos\theta)}{(\varepsilon+i\eta)^{2}-v_{F}^{2}k^{2}}=-\frac{\varepsilon}{4\pi^{2}v_{F}^{2}}\int_{0}^{\infty}\frac{k^{2}dk}{k^{2}-(\varepsilon+i\eta)^{2}/v_{F}^{2}}\int_{0}^{\pi}\sin\theta\exp(ikR\cos\theta)d\theta,
\end{equation}
where $\theta$ is the angle between vector $\mathbf{k}$ and vector $\mathbf{R}$. Integrating over $\theta$ leads to
\[
G_{0}(R;\varepsilon)=-\frac{\varepsilon}{i4\pi^{2}v_{F}^{2}R}\int_{0}^{\infty}\frac{k\left[\exp\left(ikR\right)-\exp\left(-ikR\right)\right]dk}{k^{2}-(\varepsilon+i\eta)^{2}/v_{F}^{2}}.
\]
Due to the symmetry of the integrand of $k$, the integration of $G_{0}(R;\varepsilon)$ can be recast as
\begin{equation}
G_{0}(R;\varepsilon)=-\frac{\varepsilon}{i4\pi^{2}v_{F}^{2}R}\int_{-\infty}^{\infty}\frac{k\exp\left(ikR\right)dk}{k^{2}-(\varepsilon+i\eta)^{2}/v_{F}^{2}}.
\end{equation}
After carrying out the contour integration, we arrive at the result,
\begin{equation}
G_{0}(R;\varepsilon)=-\frac{1}{4\pi v_{F}R^{2}}\varepsilon R/v_{F}\exp\left(i\varepsilon R/v_{F}\right)=-\frac{1}{4\pi v_{F}R^{2}}\xi_{\varepsilon}\exp\left(i\xi_{\varepsilon}\right)\equiv G_{0}(\xi_{\varepsilon}),\label{eq:G0R}
\end{equation}
where $\xi_{\varepsilon}=\varepsilon R/v_{F}$ is a dimensionless quantity.  
Since the latter one $G_{R}(R;\varepsilon)$ satisfies the relation
\begin{equation}
G_{R}(R;\varepsilon)=v_{F}\int\frac{d^{3}\mathbf{k}}{(2\pi)^{3}}\frac{\mathbf{k}\cdot\boldsymbol{e}_{R}\exp(i\mathbf{k}\cdot\mathbf{R})}{(\varepsilon+i\eta)^{2}-v_{F}^{2}k^{2}}=v_{F}\int\frac{d^{3}\mathbf{k}}{(2\pi)^{3}}\frac{k\cos\theta\exp(ikR\cos\theta)}{(\varepsilon+i\eta)^{2}-v_{F}^{2}k^{2}}=-i\frac{v_{F}}{\varepsilon}\frac{\partial}{\partial R}G_{0}(R;\varepsilon),
\end{equation}
we immediately get the latter
\begin{equation}
G_{R}(R;\varepsilon)=-\frac{1}{4\pi v_{F}R^{2}}i\left(1-i\varepsilon R/v_{F}\right)\exp\left(i\varepsilon R/v_{F}\right)=-\frac{1}{4\pi v_{F}R^{2}}i\left(1-i\xi_{\varepsilon}\right)\exp\left(i\xi_{\varepsilon}\right)\equiv G_{R}(\xi_{\varepsilon}).\label{eq:GRR}
\end{equation}
Note that the procedure above is also applicable for the Weyl SMs. Taking into account the separation
of Weyl nodes in the internode process, we get the corresponding Green function for Weyl SMs without TRS in energy-coordinate representation
\begin{equation}
G_{\chi}\left(\text{\ensuremath{\mathbf{\pm R}},\ensuremath{\varepsilon}}\right)=\int\frac{d^{3}\mathbf{k}}{\left(2\pi\right)^{3}}G_{\chi}\left(\mathbf{k},\varepsilon\right)\exp\bigl(\pm i\left(\mathbf{k}+\chi\mathbf{Q}\right)\cdot\mathbf{R}\bigr)=\exp\bigl(\pm i\chi\mathbf{Q}\cdot\mathbf{R}\bigr)\int\frac{d^{3}\mathbf{k}}{\left(2\pi\right)^{3}}G_{\chi}\left(\mathbf{k},\varepsilon\right)\exp\bigl(\pm i\mathbf{k}\cdot\mathbf{R}\bigr),
\end{equation}
which is the counterpart of the Dirac SMs multiplied by a factor of $\exp\left(\pm i\chi\mathbf{Q}\cdot\mathbf{R}\right)$.
For noncentrosymmetric Weyl SMs, the corresponding Green function in the energy-coordinate representation
can be directly obtained from that of Dirac SMs by replacing $\varepsilon$ with $\varepsilon-\chi\mathrm{Q}_{0}$.

\section{Range functions and their asymptotic behaviors}

To simplify the presentation, we only show the detailed calculation of the RKKY interaction of isotropic Dirac SMs.
Substituting Eq.~$(\ref{eq:GchiR})$ into Eq.~$\left(\ref{eq:Fchialpha1}\right)$, we get the corresponding RKKY interaction
\begin{eqnarray}
 &  & H_{\mathrm{RKKY}}^{\mathrm{D}}(R;\varepsilon_{F})=\mathrm{Im}\sum_{\chi,\chi^{\prime},\alpha,\beta}\Big[J^{2}\delta_{\chi\chi^{\prime}}+\lambda^{2}(1-\delta_{\chi\chi^{\prime}})\Big]\mathcal{F}_{\alpha\beta}^{\chi\chi^{\prime}}(R;\varepsilon_{F})\boldsymbol{S}_{1}^{\alpha} \boldsymbol{S}_{2}^{\beta},\label{eq:rkkydefDir}
\end{eqnarray}
with the functions
\begin{eqnarray}
 &  & \mathcal{F}_{\alpha\beta}^{\chi\chi^{\prime}}(R;\varepsilon_{F})
 =-\frac{1}{\pi}\int_{-\infty}^{\varepsilon_{F}}d\varepsilon\,\mathrm{Tr}
 \left[\sigma_{\alpha}\Big(\sigma_{0}G_{0}(R;\varepsilon)+\chi\sigma_{j}G_{R}(R;\varepsilon)\Big)
 \sigma_{\beta}\Big(\sigma_{0}G_{0}(R;\varepsilon)-\chi^{\prime}\sigma_{j}
 G_{R}(R;\varepsilon)\Big)\right],\label{eq:FchialphaDir}
\end{eqnarray}
where we have assumed $\boldsymbol{e}_{R}=\boldsymbol{e}_{j}$, which
implies $\boldsymbol{\sigma}\cdot\boldsymbol{e}_{R}=\sigma_{j}$. For the remainder of this section,
we compute the RKKY interactions for the cases with and without spin-momentum locking, respectively.

\subsection{Case with spin-momentum locking}

For the case with spin-momentum locking, the trace term in Eq.~$(\ref{eq:FchialphaDir})$ can be calculated as follows,
\begin{align}
 & \mathrm{Tr}\left[\sigma_{\alpha}\Big(\sigma_{0}G_{0}(R;\varepsilon)+\chi\sigma_{j}G_{R}(R;\varepsilon)\Big)\sigma_{\beta}\Big(\sigma_{0}G_{0}(R;\varepsilon)-\chi^{\prime}\sigma_{j}G_{R}(R;\varepsilon)\Big)\right]\nonumber \\
= & G_{0}(R;\varepsilon)G_{0}(R;\varepsilon)\mathrm{Tr}\left(\sigma_{\alpha}\sigma_{0}\sigma_{\beta}\sigma_{0}\right)-\chi\chi^{\prime}G_{R}(R;\varepsilon)G_{R}(R;\varepsilon)\mathrm{Tr}\left(\sigma_{\alpha}\sigma_{j}\sigma_{\beta}\sigma_{j}\right)\nonumber \\
 & +\chi G_{R}(R;\varepsilon)G_{0}(R;\varepsilon)\mathrm{Tr}\left(\sigma_{\alpha}\sigma_{j}\sigma_{\beta}\sigma_{0}\right)-\chi^{\prime}G_{0}(R;\varepsilon)G_{R}(R;\varepsilon)\mathrm{Tr}\left(\sigma_{\alpha}\sigma_{0}\sigma_{\beta}\sigma_{j}\right).
\end{align}
Substituting it into Eq.~$\left(\ref{eq:FchialphaDir}\right)$, we have the functions
\begin{align*}
\mathcal{F}_{\alpha\beta}^{\chi\chi^{\prime}}(R;\varepsilon_{F})=-\frac{1}{\pi}\left(\frac{v_{F}}{R}\right)\int_{-\infty}^{\xi_{F}}d\xi\,\Big[ & 2\delta_{\alpha\beta}\Big(G_{0}(\xi)G_{0}(\xi)+\chi\chi^{\prime}G_{R}(\xi)G_{R}(\xi)\Big)\\
 & -4\delta_{\alpha j}\delta_{\beta j}\chi\chi^{\prime}G_{R}(\xi)G_{R}(\xi)-2i\varepsilon_{\alpha\beta j}(\chi+\chi^{\prime})G_{0}(\xi)G_{R}(\xi)\Big],
\end{align*}
with $\xi_{F}=\varepsilon_{F}R/v_{F}$. Making use of Eq.~$\left(\ref{eq:rkkydefDir}\right)$
and summing over $\chi$, $\chi^{\prime}$, $\alpha$, and $\beta$, we get the RKKY interaction of the Dirac SMs
\begin{eqnarray}
 &  & H_{\mathrm{RKKY}}^{\mathrm{D}}(R;\varepsilon_{F})=F_{\mathrm{H}}^{\mathrm{D}}(\xi_{F}) \boldsymbol{S}_{1}\cdot \boldsymbol{S}_{2}+F_{\mathrm{Ising}}^{\mathrm{D}}(\xi_{F})\boldsymbol{S}_{1}^{j} \boldsymbol{S}_{2}^{j},
\end{eqnarray}
where the range functions are given as
\begin{align*}
F_{\mathrm{H}}^{\mathrm{D}}(\xi_{F}) & =-\frac{1}{\pi}\left(\frac{v_{F}}{R}\right)\mathrm{Im}\int_{-\infty}^{\xi_{F}}d\xi\,4\left[J^{2}\Big(G_{0}(\xi)G_{0}(\xi)+G_{R}(\xi)G_{R}(\xi)\Big)+\lambda^{2}\Big(G_{0}(\xi)G_{0}(\xi)-G_{R}(\xi)G_{R}(\xi)\Big)\right]\\
 & =-\frac{1}{\pi}\left(\frac{v_{F}}{R}\right)\left(-\frac{1}{4\pi v_{F}R^{2}}\right)^{2}4\,\mathrm{Im}\Bigl[J^{2}\Bigl(2\mathcal{I}_{2}(\xi_{F},2)+2i\mathcal{I}_{1}(\xi_{F},2)-\mathcal{I}_{0}(\xi_{F},2)\Bigr)+\lambda^{2}\Bigl(-2i\mathcal{I}_{1}(\xi_{F},2)+\mathcal{I}_{0}(\xi_{F},2)\Bigr)\Bigr]\\
F_{\mathrm{Ising}}^{\mathrm{D}}(\xi_{F}) & =-\frac{1}{\pi}\left(\frac{v_{F}}{R}\right)\mathrm{Im}\int_{-\infty}^{\xi_{F}}d\xi\,8\left(\lambda^{2}-J^{2}\right)G_{R}(\xi)G_{R}(\xi)\\
 & =-\frac{1}{\pi}\left(\frac{v_{F}}{R}\right)\left(-\frac{1}{4\pi v_{F}R^{2}}\right)^{2}8\left(\lambda^{2}-J^{2}\right)\mathrm{Im}\Bigl[\mathcal{I}_{2}(\xi_{F},2)+2i\mathcal{I}_{1}(\xi_{F},2)-\mathcal{I}_{0}(\xi_{F},2)\Bigr],
\end{align*}
where we have utilized Eqs.~$(\ref{eq:G0R})$ and $(\ref{eq:GRR})$.
Making use of the basic integral $\mathcal{I}_{n}(\xi_{F},a)$ given in Sec.~\ref{sec:BIs}, we finally obtain
\begin{align}
F_{\mathrm{H}}^{\mathrm{D}}(\xi_{F}) & =-\Bigl[J^{2}\Bigl((3-2\xi_{F}^{2})\cos(2\xi_{F})+4\xi_{F}\sin(2\xi_{F})\Bigr)-2\lambda^{2}\Bigl(\cos(2\xi_{F})+\xi_{F}\sin(2\xi_{F})\Bigr)\Bigr]\Bigl/8\pi^{3}v_{F}R^{5},\\
F_{\mathrm{Ising}}^{\mathrm{D}}(\xi_{F}) & =-\left(J^{2}-\lambda^{2}\right)\Bigl[(2\xi_{F}^{2}-5)\cos(2\xi_{F})-6\xi_{F}\sin(2\xi_{F})\Bigr]\Bigl/8\pi^{3}v_{F}R^{5}.
\end{align}
Note that the Ising term $F_{\mathrm{Ising}}^{\mathrm{D}}(\xi_{F})$
vanishes for homogeneous coupling case $\lambda=J$. The asymptotic behavior of
range function $F_{\mathrm{H}}^{\mathrm{D}}(\xi_{F})$
at long range ($\xi_{F}\gg1$) reduces to $F_{\mathrm{H}}^{\mathrm{D}}(\xi_{F})\approx J^{2}\xi_{F}^{2}\cos(2\xi_{F})\Bigl/4\pi^{3}v_{F}R^{5}
=J^{2}\varepsilon_{F}^{2}\cos(2\xi_{F})\Bigl/4\pi^{3}v_{F}^{3}R^{3}$.

The procedure above is also applicable for the Weyl SMs.
For the Weyl SMs without TRS, $\mathcal{F}_{\alpha\beta}^{\chi\chi^{\prime}}(R;\varepsilon_{F})$
in Eq.~$\left(\ref{eq:FchialphaDir}\right)$ is substituted by $\exp\left[i(\chi-\chi^{\prime})\mathbf{Q}\cdot\mathbf{R}\right]\mathcal{F}_{\alpha\beta}^{\chi\chi^{\prime}}(R;\varepsilon_{F})$.
After summing over $\chi$, $\chi^{\prime}$, $\alpha$, and $\beta$, the corresponding RKKY interaction becomes
\begin{eqnarray}
 &  & H_{\mathrm{RKKY}}^{\mathrm{I}}(R;\varepsilon_{F})=F_{\mathrm{H}}^{\mathrm{I}}(\xi_{F}) \boldsymbol{S}_{1}\cdot \boldsymbol{S}_{2}+F_{\mathrm{Ising}}^{\mathrm{I}}(\xi_{F})\mathrm{S}_{1}^{j}\mathrm{S}_{2}^{j},
\end{eqnarray}
where the range functions are given by
\begin{align}
F_{\mathrm{H}}^{\mathrm{I}}(\xi_{F}) & =-\Bigl[J^{2}\Bigl((3-2\xi_{F}^{2})\cos(2\xi_{F})+4\xi_{F}\sin(2\xi_{F})\Bigr)-2\lambda^{2}\cos\left(2\mathbf{Q}\cdot\mathbf{R}\right)\Bigl(\cos(2\xi_{F})+\xi_{F}\sin(2\xi_{F})\Bigr)\Bigr]\Bigl/8\pi^{3}v_{F}R^{5},\\
F_{\mathrm{Ising}}^{\mathrm{I}}(\xi_{F}) & =-\Bigl(J^{2}-\lambda^{2}\cos\left(2\mathbf{Q}\cdot\mathbf{R}\right)\Bigr)\Bigl[(2\xi_{F}^{2}-5)\cos(2\xi_{F})-6\xi_{F}\sin(2\xi_{F})\Bigr]\Bigl/8\pi^{3}v_{F}R^{5}.
\end{align}
Two remarks on the range functions are in order here. First, $F_{\mathrm{H}}^{\mathrm{I}}(\xi_{F})$ and $F_{\mathrm{Ising}}^{\mathrm{I}}(\xi_{F})$
can be obtained from $F_{\mathrm{H}}^{\mathrm{D}}(\xi_{F})$ and $F_{\mathrm{Ising}}^{\mathrm{D}}(\xi_{F})$ by replacing
$\lambda^{2}$ by $\lambda^{2}\cos\left(2\mathbf{Q}\cdot\mathbf{R}\right)$.
Second, the internode process gives rise to an additional oscillating term proportional to $\cos\left(2\mathbf{Q}\cdot\mathbf{R}\right)$.
Because of its long range and oscillatory nature of the RKKY
interaction, for a large momentum separation $2 \mathbf{Q}$, the part from the internode process will vanish after averaging over position.
Hence, the intranode process dominates the RKKY interaction in the Weyl SMs without TRS, leading to a non-vanishing Ising term.
Therefore the range functions can be expressed as
\begin{align}
F_{\mathrm{H}}^{\mathrm{I}}(\xi_{F}) & =-J^{2}\Bigl((3-2\xi_{F}^{2})\cos(2\xi_{F})+4\xi_{F}\sin(2\xi_{F})\Bigr)\Bigl/8\pi^{3}v_{F}R^{5},\\
F_{\mathrm{Ising}}^{\mathrm{I}}(\xi_{F}) & =-J^{2}\Bigl[(2\xi_{F}^{2}-5)\cos(2\xi_{F})-6\xi_{F}\sin(2\xi_{F})\Bigr]\Bigl/8\pi^{3}v_{F}R^{5}.
\end{align}
The asymptotic range functions at long range ($\xi_{F}\gg1$) become
\begin{align}
F_{\mathrm{H}}^{\mathrm{I}}(\xi_{F}) & \approx J^{2}\xi_{F}^{2}\cos(2\xi_{F})\Bigl/4\pi^{3}v_{F}R^{5}=J^{2}\varepsilon_{F}^{2}\cos(2\xi_{F})\Bigl/4\pi^{3}v_{F}^{3}R^{3},\\
F_{\mathrm{Ising}}^{\mathrm{I}}(\xi_{F}) & \approx-J^{2}\xi_{F}^{2}\cos(2\xi_{F})\Bigl/4\pi^{3}v_{F}R^{5}=-J^{2}\varepsilon_{F}^{2}\cos(2\xi_{F})\Bigl/4\pi^{3}v_{F}^{3}R^{3}.
\end{align}
For the noncentrosymmetric Weyl SMs, $\mathcal{F}_{\alpha\beta}^{\chi\chi^{\prime}}(R;\varepsilon_{F})$ in Eq.~$(\ref{eq:FchialphaDir})$ turns out to be
\begin{align}
\mathcal{F}_{\alpha\beta}^{\chi\chi^{\prime}}(R;\varepsilon_{F})
=-\frac{1}{\pi}\int_{-\infty}^{\varepsilon_{F}}d\varepsilon\,\mathrm{Tr}\Bigl[ & \sigma_{\alpha}\Big(\sigma_{0}G_{0}(R;\varepsilon-\chi\mathrm{Q_{0}})
+\chi\sigma_{j}G_{R}(R;\varepsilon-\chi\mathrm{Q_{0}})\Big)\\
\times & \sigma_{\beta}\Big(\sigma_{0}G_{0}(R;\varepsilon-\chi^{\prime}\mathrm{Q_{0}})
-\chi^{\prime}\sigma_{j}G_{R}(R;\varepsilon-\chi^{\prime}\mathrm{Q_{0}})\Big)\Bigr].
\end{align}
After summing over the indices for spin and chirality, the corresponding RKKY interaction becomes
\begin{eqnarray}
 &  & H_{\mathrm{RKKY}}^{\mathrm{TR}}(R;\varepsilon_{F})=F_{\mathrm{H}}^{\mathrm{TR}}(\xi_{F,}\zeta) \boldsymbol{S}_{1}\cdot \boldsymbol{S}_{2}+F_{\mathrm{Ising}}^{\mathrm{TR}}(\xi_{F},\zeta)\mathrm{S}_{1}^{j}\mathrm{S}_{2}^{j}+F_{\mathrm{DM}}^{\mathrm{TR}}(\xi_{F},\zeta)\left(\boldsymbol{S}_{1}\times \boldsymbol{S}_{2}\right)_{j},
\end{eqnarray}
where $\zeta=\mathrm{Q_{0}}R/v_{F}$ and the three range functions are given as
\begin{align}
F_{\mathrm{H}}^{\mathrm{TR}}(\xi_{F,}\zeta)=-\Bigl[J^{2} & \Bigl((3-2\xi_{F}^{2}-2\zeta^{2})\cos(2\xi_{F})\cos\left(2\zeta\right)+4\xi_{F}\zeta\sin(2\xi_{F})\sin\left(2\zeta\right)+4\xi_{F}\sin(2\xi_{F})\cos\left(2\zeta\right)\nonumber \\
 & +4\zeta\cos(2\xi_{F})\sin\left(2\zeta\right)\Bigr)-2\lambda^{2}\Bigl(\cos(2\xi_{F})+\xi_{F}\sin(2\xi_{F})\Bigr)\Bigr]\Bigr/8\pi^{3}v_{F}R^{5},\\
F_{\mathrm{Ising}}^{\mathrm{TR}}(\xi_{F,}\zeta)=-\Bigl[J^{2} & \Bigl((2\xi_{F}^{2}+2\zeta^{2}-5)\cos(2\xi_{F})\cos\left(2\zeta\right)-4\xi_{F}\zeta\sin(2\xi_{F})\sin\left(2\zeta\right)-6\xi_{F}\sin(2\xi_{F})\cos\left(2\zeta\right)\nonumber \\
 & -6\zeta\cos(2\xi_{F})\sin\left(2\zeta\right)\Bigr)+\lambda^{2}\Bigl((5-2\xi_{F}^{2}+2\zeta^{2})\cos(2\xi_{F})+6\xi_{F}\sin(2\xi_{F})\Bigr)\Bigr]\Bigr/8\pi^{3}v_{F}R^{5},\\
F_{\mathrm{DM}}^{\mathrm{TR}}(\xi_{F,}\zeta)=-\Bigl[J^{2} & \Bigl((2\xi_{F}^{2}+2\zeta^{2}-2)\cos(2\xi_{F})\sin\left(2\zeta\right)+4\xi_{F}\zeta\sin(2\xi_{F})\cos\left(2\zeta\right)-4\xi_{F}\sin(2\xi_{F})\sin\left(2\zeta\right)\nonumber \\
 & +4\zeta\cos(2\xi_{F})\cos\left(2\zeta\right)\Bigr)-2\lambda^{2}\zeta\cos(2\xi_{F})\Bigr]\Bigr/8\pi^{3}v_{F}R^{5}.
\end{align}
It is clear that for $\zeta=0$, i.e., the Dirac SMs case, the DM term $F_{\mathrm{DM}}^{\mathrm{TR}}(\xi_{F,}0)$ vanishes,
and the range functions $F_{\mathrm{H}}^{\mathrm{TR}}(\xi_{F,}\zeta)$
and $F_{\mathrm{Ising}}^{\mathrm{TR}}(\xi_{F,}\zeta)$ reduce to $F_{\mathrm{H}}^{\mathrm{D}}(\xi_{F})$
and $F_{\mathrm{Ising}}^{\mathrm{D}}(\xi_{F})$, respectively. In addition,
the range functions of RKKY interaction for the noncentrosymmetric Weyl SMs oscillate with two distinct periods and form a beating pattern.
Note that these behaviors can be roughly described by the asymptotic expressions
at long range. For $\xi_{F}\gg\zeta$ and $\xi_{F}\gg1$, the asymptotic range functions become
\begin{align*}
F_{\mathrm{H}}^{\mathrm{TR}}(\xi_{F,}\zeta) & \approx J^{2}\xi_{F}^{2}\cos(2\xi_{F})\cos\left(2\zeta\right)\Bigr/4\pi^{3}v_{F}R^{5},\\
F_{\mathrm{Ising}}^{\mathrm{TR}}(\xi_{F,}\zeta) & \approx-\xi_{F}^{2}\cos(2\xi_{F})\Bigl(J^{2}\cos\left(2\zeta\right)-\lambda^{2}\Bigr)\Bigr/4\pi^{3}v_{F}R^{5},\\
F_{\mathrm{DM}}^{\mathrm{TR}}(\xi_{F,}\zeta) & \approx-J^{2}\xi_{F}^{2}\cos(2\xi_{F})\sin\left(2\zeta\right)\Bigr/4\pi^{3}v_{F}R^{5}.
\end{align*}
For $\zeta\gg\xi_{F}$ and $\zeta \gg 1$, the range functions read
\begin{align*}
F_{\mathrm{H}}^{\mathrm{TR}}(\xi_{F,}\zeta) & \approx J^{2}\zeta^{2}\cos(2\xi_{F})\cos\left(2\zeta\right)\Bigr/4\pi^{3}v_{F}R^{5},\\
F_{\mathrm{Ising}}^{\mathrm{TR}}(\xi_{F,}\zeta) & \approx-\zeta^{2}\cos(2\xi_{F})\Bigl(J^{2}\cos\left(2\zeta\right)+\lambda^{2}\Bigr)\Bigr/4\pi^{3}v_{F}R^{5},\\
F_{\mathrm{DM}}^{\mathrm{TR}}(\xi_{F,}\zeta) & \approx-J^{2}\zeta^{2}\cos(2\xi_{F})\sin\left(2\zeta\right)\Bigr/4\pi^{3}v_{F}R^{5}.
\end{align*}
It should be emphasized that these two different asymptotic range functions account for the distinct beating pattens in Fig.~2(a) and 2(b) in the main text, respectively.

\subsection{Case without spin-momentum locking}
%
For the case without spin-momentum locking, the trace term in $\mathcal{F}_{\alpha\beta}^{\chi\chi^{\prime}}(R;\varepsilon_{F})$ reads
\begin{align}
 &\mathrm{Tr}\left[\sigma_{\alpha}\Big( \tilde \tau_{0} G_{0}(R;\varepsilon)
 +\chi \tilde \tau_{j}G_{R}(R;\varepsilon)\Big)\sigma_{\beta}
 \Big(\tilde \tau_{0}G_{0}(R;\varepsilon)-\chi^{\prime} \tilde \tau_{j} G_{R}(R;\varepsilon)\Big)\right]\nonumber\\
=& G_{0}(R;\varepsilon)G_{0}(R;\varepsilon)\mathrm{Tr}
 (\sigma_{\alpha}\sigma_{\beta})\mathrm{Tr}
 ( \tilde \tau_{0}\tau_{0})-\chi\chi^{\prime}G_{R}(R;\varepsilon)
 G_{R}(R;\varepsilon)\mathrm{Tr}(\sigma_{\alpha}\sigma_{\beta})
 \mathrm{Tr}( \tilde \tau_{j} \tilde \tau_{j})\nonumber\\
 & +\chi G_{R}(R;\varepsilon)G_{0}(R;\varepsilon)
 \mathrm{Tr}(\sigma_{\alpha}\sigma_{\beta})
 \mathrm{Tr} ( \tilde \tau_{j} \tilde \tau_{0})-\chi^{\prime}G_{0}(R;\varepsilon)
 G_{R}(R;\varepsilon)\mathrm{Tr}(\sigma_{\alpha}\sigma_{\beta})
 \mathrm{Tr} ( \tilde \tau_{0} \tilde \tau_{j}),
\end{align}
Where the identity matrix $\tilde \tau_0$ and Pauli matrix $\tilde \tau_i$ here refer to other pseudospin degree of freedom rather than the chirality.  
Substituting the above trace term into the definition of $\mathcal{F}_{\alpha\beta}^{\chi\chi^{\prime}}(R;\varepsilon_{F})$, we have
\begin{eqnarray}
 &  & \mathcal{F}_{\alpha\beta}^{\chi\chi^{\prime}}(R;\varepsilon_{F})=-\frac{1}{\pi}\left(\frac{v_{F}}{R}\right)\int_{-\infty}^{\xi_{F}}d\xi\,4\delta_{\alpha\beta}\Big(G_{0}(\xi)G_{0}(\xi)-\chi\chi^{\prime}G_{R}(\xi)G_{R}(\xi)\Big).
\end{eqnarray}
After summing over the indices for the spin and chirality degree of freedoms, we get the corresponding RKKY interaction
\begin{eqnarray}
 &  & H_{\mathrm{RKKY}}^{\mathrm{D}}(R;\varepsilon_{F})=\tilde{F}_{\mathrm{H}}^{\mathrm{D}}(\xi_{F}) \boldsymbol{S}_{1}\cdot \boldsymbol{S}_{2},
\end{eqnarray}
where the tilde is used to distinguish the range function of the case without spin-momentum locking from that with spin-momentum locking.
Utilizing the explicit expressions of basic integral $\mathcal{I}_{n}(\xi_{F},a)$ evaluated in Sec.~\ref{sec:BIs}, we finally obtain the range function,
\begin{eqnarray}
\tilde{F}_{\mathrm{H}}^{\mathrm{D}}(\xi_{F})
=\Bigl[4J^{2}\Bigl(\cos(2\xi_{F})+\xi_{F}\sin(2\xi_{F})\Bigr)
-\lambda^{2}\Bigl(2(3-2\xi_{F}^{2})\cos(2\xi_{F})+8\xi_{F}\sin(2\xi_{F})\Bigr)\Bigr]\Bigr/8\pi^{3}v_{F}R^{5}.
\end{eqnarray}
Similarly, the RKKY interaction for the TRS-breaking Weyl SMs reads
\begin{eqnarray}
H_{\mathrm{RKKY}}^{\mathrm{I}}(R;\varepsilon_{F})=\tilde{F}_{\mathrm{H}}^{\mathrm{I}}(\xi_{F}) \boldsymbol{S}_{1}\cdot \boldsymbol{S}_{2},
\end{eqnarray}
where the range function is
\begin{eqnarray}
\tilde{F}_{\mathrm{H}}^{\mathrm{I}}(\xi_{F})
=\Bigl[4J^{2}\Bigl(\cos(2\xi_{F})+\xi_{F}\sin(2\xi_{F})\Bigr)
-\lambda^{2}\cos\left(2\mathbf{Q}\cdot\mathbf{R}\right)
\Bigl(2(3-2\xi_{F}^{2})\cos(2\xi_{F})+8\xi_{F}\sin(2\xi_{F})\Bigr)\Bigr]\Bigr/8\pi^{3}v_{F}R^{5},
\end{eqnarray}
and that for the noncentrosymmetric Weyl SMs,
\begin{eqnarray}
 &  & H_{\mathrm{RKKY}}^{\mathrm{TR}}(R;\varepsilon_{F})=\tilde{F}_{\mathrm{H}}^{\mathrm{TR}}(\xi_{F}) \boldsymbol{S}_{1}\cdot \boldsymbol{S}_{2},
\end{eqnarray}
where the range function has the form
\begin{align}
\tilde{F}_{\mathrm{H}}^{\mathrm{TR}}(\xi_{F})=\Bigl[& 4J^{2}\Bigl(\cos(2\xi_{F})\cos(2\zeta)+\xi_{F}\sin(2\xi_{F})\cos(2\zeta)
+\zeta\cos(2\xi_{F})\sin(2\zeta)\Bigr)\nonumber \\
 & -\lambda^{2}\Bigl(2(3-2\xi_{F}^{2}+2\zeta^{2})\cos(2\xi_{F})+8\xi_{F}\sin(2\xi_{F})\Bigr)\Bigr]
 \Bigr/8\pi^{3}v_{F}R^{5}.
\end{align}
%

\section{Effect of anisotropy}
\label{SecIV}
%
Let us turn to consider the RKKY interaction of the anisotropic Dirac SMs.
The effective Hamiltonian for anisotropic Dirac SMs in the vicinity
of the Weyl node of chirality $\chi$ takes the form
\begin{eqnarray}
\tilde{H}_{0\lambda}^{A}=\chi\left(v_{x}\tilde{k}_{x}\sigma_{x}+v_{y}\tilde{k}_{y}\sigma_{y}+v_{z}\tilde{k}_{z}\sigma_{z}\right),\label{H0an}
\label{HA}
\end{eqnarray}
where $v_{\alpha}$ is the Fermi velocity in the $\alpha$-direction
with $\alpha = x,y,z$, which generally satisfies the relation $\left|v_{x}\right|\neq\left|v_{y}\right|\neq\left|v_{z}\right|$.
The superscript $A$ stands for the anisotropy of Weyl node.
To simplify following discussion, we set
$v_{\alpha}>0$. The wave vector $\tilde{k}_{\alpha}$ denotes the
derivation from anisotropic Weyl node $\chi$.
The energy dispersion of Eq.~(\ref{HA}) is given as
$\mathscr{\varepsilon}_{\tilde{k}}=\pm\sqrt{v_{x}^{2}\tilde{k}_{x}^{2}+v_{y}^{2}\tilde{k}_{y}^{2}+v_{z}^{2}\tilde{k}_{z}^{2}}$.
The Green function in momentum space is
$G_{\chi}^{-1}(\tilde{\mathbf{k}};\varepsilon) = (\varepsilon+i\eta) \sigma_{0}-\tilde{H}_{0\lambda}^{A}$.
It is instructive to introduce the following associated transformation as
\begin{eqnarray}
\Bigl(v_{\alpha}\tilde{Q}_{\alpha},v_{\alpha}\tilde{k}_{\alpha},v_{F}\tilde{R}_{\alpha}\Bigr)=\Bigl(v_{F}Q_{\alpha},v_{F}k_{\alpha},v_{\alpha}R_{\alpha}\Bigr),
\end{eqnarray}
which leads to two consequences. First, it relates the Green function of an anisotropic Weyl node to that of the isotropic one as
\[
G_{\chi}(\pm\tilde{\mathbf{R}};\varepsilon)=\int\frac{d^{3}\tilde{\mathbf{k}}}{(2\pi)^{3}}G_{\chi}(\tilde{\mathbf{k}};\varepsilon)\exp\left(\pm i\tilde{\mathbf{k}}\cdot\tilde{\mathbf{R}}\right)=\lambda_{A}\int\frac{d^{3}\mathbf{k}}{(2\pi)^{3}}G_{\chi}(\mathbf{k};\varepsilon)\exp\left(\pm i\mathbf{k}\cdot\mathbf{R}\right)=\lambda_{A}G(\pm\mathbf{R};\varepsilon),
\]
and
\begin{equation}
\mathcal{G\left(\text{\ensuremath{\pm\mathbf{\tilde{\mathbf{R}}}},\ensuremath{\varepsilon}}\right)}=\begin{pmatrix}G_{+}\mbox{(\ensuremath{\pm\tilde{\mathbf{R}}},\ensuremath{\varepsilon})} & 0\\
0 & G_{-}\mbox{(\ensuremath{\pm\tilde{\mathbf{R}}},\ensuremath{\varepsilon})}
\end{pmatrix}=\lambda_{A}\,\begin{pmatrix}G_{+}\mbox{(\ensuremath{\pm\text{\ensuremath{\mathbf{R}},\ensuremath{\varepsilon}}})} & 0\\
0 & G_{-}\mbox{(\ensuremath{\pm\text{\ensuremath{\mathbf{R}},\ensuremath{\varepsilon}}})}
\end{pmatrix}=\lambda_{A}\,\mathcal{G\left(\text{\ensuremath{\pm\mathbf{R}},\ensuremath{\varepsilon}}\right)}
\end{equation}
where $\lambda_{A}=v_{F}^{3}/\bigl(v_{x}v_{y}v_{z}\bigr)$ measures the anisotropy.
Second, it converts a Fermi elliptic sphere to a Fermi sphere but preserves its volume, namely,
\[
\mathscr{\varepsilon}_{\tilde{k}}=\pm\sqrt{v_{x}^{2}\tilde{k}_{x}^{2}+v_{y}^{2}\tilde{k}_{y}^{2}+v_{z}^{2}\tilde{k}_{z}^{2}}=\pm v_{F}\sqrt{k_{x}^{2}+k_{y}^{2}+k_{z}^{2}}=\mathscr{\varepsilon}_{k}.
\]
We choose $\tilde{\mathbf{R}}$ to be aligned to the $j$-direction ($\boldsymbol{e}_{R}=\boldsymbol{e}_{j}$) and have
\[
\tilde{\mathbf{R}}=\tilde{R}\boldsymbol{e}_{j}
=\frac{v_{j}}{v_{F}}R\boldsymbol{e}_{j}=\frac{v_{j}}{v_{F}}\mathbf{R}.
\]
Substituting this result into the original expression of the RKKY interaction leads to
\begin{align}
\tilde{H}_{\mathrm{RKKY}}(\tilde{R}) & =-\frac{1}{\pi}\mathrm{Im}\int_{-\infty}^{\varepsilon_{F}}d\varepsilon\,\mathrm{Tr}
\Bigl[\left(J \tau_{0}+\lambda \tau_{x}\right)\boldsymbol{S}_{1}\cdot\boldsymbol{\sigma}\,
\mathcal{G}\left(\tilde{\mathbf{R}},\varepsilon\right)\left(J \tau_{0}+\lambda \tau_{x}\right)
\boldsymbol{S}_{2}\cdot\boldsymbol{\sigma}\,\mathcal{G}\left(-\tilde{\mathbf{R}},\varepsilon\right)\Bigr]\nonumber \\
 & =\lambda_{A}^{2}\, H_{\mathrm{RKKY}}(R)=\lambda_{A}^{2}\, H_{\mathrm{RKKY}}\Bigl(\frac{v_{F}}{v_{j}}\tilde{R}\Bigr),  
\end{align}
which converts the expression of RKKY interaction of anisotropic Dirac SMs to the counterpart for isotropic Dirac SMs.
Compared with the isotropic case, the anisotropy not only rescales the amplitude of RKKY interaction by $\lambda_{A}^{2}$,
but also the effective spatial separation between two impurities by $v_{F}/v_{j}$. Note that this procedure can also be applied to the anisotropic Weyl SMs.

%

\section{Basic integrals}
\label{sec:BIs}
In this section, we evaluate the following basic integrals
\begin{eqnarray}
\mathcal{I}_{n}(\xi_{F},a)=\int_{-\infty}^{\xi_{F}}\xi^{n}\exp(ia\xi)d\xi,~~~~~~n=0,1,2.
\end{eqnarray}
We at first consider the integral
\begin{eqnarray}
\mathcal{I}_{0}(\xi_{F},a)=\int_{-\infty}^{\xi_{F}}\exp(ia\xi)d\xi,\label{eq:arguement}
\end{eqnarray}
which does not converge if $a$ is real. To carry out this integral, we
make use of the trick in Ref.~\cite{Abrikosov} by adding a negative infinitesimal
imaginary part to $a$, i.e., $a\to a-i\eta$ with $\eta>0$,
\begin{eqnarray}
\mathcal{I}_{0}(\xi_{F},a)\to\int_{-\infty}^{\xi_{F}}\exp[i(a-i\eta)\xi]d\xi=\frac{\exp(ia\xi)\exp(\eta\xi)}{i(a-i\eta)}\bigg|_{-\infty}^{\xi_{F}}=\frac{-i\exp(ia\xi_{F})\exp(\eta\xi_{F})}{a-i\eta}.
\end{eqnarray}
After taking the limit $\eta\to0^{+}$, we have
\begin{equation}
\mathcal{I}_{0}(\xi_{F},a)=\lim_{\eta\to0^{+}}\frac{-i\exp(ia\xi_{F})\exp(\eta\xi_{F})}{a-i\eta}=\frac{-i\exp(ia\xi_{F})}{a}=\frac{\sin(a\xi_{F})}{a}-i\frac{\cos(a\xi_{F})}{a}.\label{eq:form0}
\end{equation}
It is straightforward to obtain the other basic integrals
\begin{eqnarray}
 &  & \mathcal{I}_{1}(\xi_{F},a)=-i\frac{\partial}{\partial a}\mathcal{I}_{0}(\xi_{F},a)=\frac{\cos(a\xi_{F})+a\xi_{F}\sin(a\xi_{F})}{a^{2}}+i\frac{\sin(a\xi_{F})-a\xi_{F}\cos(a\xi_{F})}{a^{2}},\label{eq:form1}\\
 &  & \mathcal{I}_{2}(\xi_{F},a)=(-i)^{2}\frac{\partial^{2}}{\partial a^{2}}\mathcal{I}_{0}(\xi_{F},a)=-i\frac{\partial}{\partial a}\mathcal{I}_{1}(\xi_{F},a)\nonumber \\
 &  & \hspace{1.4cm}=\frac{(a^{2}\xi_{F}^{2}-2)\sin(a\xi_{F})+2a\xi_{F}\cos(a\xi_{F})}{a^{3}}+i\frac{(2-a^{2}\xi_{F}^{2})\cos(a\xi_{F})+2a\xi_{F}\sin(a\xi_{F})}{a^{3}}.\label{eq:form2}
\end{eqnarray}
%


\begin{thebibliography}{1}
%
\bibitem{SMYoung} S. M. Young, S. Zaheer, J. C. Y. Teo, C. L. Kane, E. J. Mele, and A. M. Rappe, Phys. Rev. Lett. {\bf 108}, 140405 (2012).

\bibitem{XGWanPRB} X. Wan, A. M. Turner, A. Vishwanath, and S. Y. Savrasov, Phys. Rev. B {\bf 83}, 205101 (2011).

\bibitem{Volovik} G. E. Volovik, \emph{The Universe in a Helium Droplet} (Clarendon, Oxford, UK, 2003).

\bibitem{XiaoRMP} D. Xiao, M.-C. Chang, and Q. Niu, Rev. Mod. Phys. {\bf 82}, 1959 (2010).

\bibitem{excitonSF} H. Wei, S.-P. Chao, and V. Aji, Phys. Rev. Lett. {\bf 109}, 196403 (2012).

\bibitem{SC} G. Y. Cho, J. H. Bardarson, Y.-M. Lu, and J. E. Moore, Phys. Rev. B {\bf 86}, 214514 (2012); S. A. Yang, H. Pan, and F. Zhang, Phys. Rev. Lett. {\bf 113}, 046401 (2014).

\bibitem{SYXUFermiArc} S.-Y. Xu, C. Liu, S. K. Kushwaha, R. Sankar, J. W. Krizan, I. Belopolski, M. Neupane, G. Bian, N. Alidoust, T.-R. Chang, H.-T. Jeng, C.-Y. Huang, W.-F. Tsai, H. Lin, P. P. Shibayev, F.-C. Chou, R. J. Cava, and M. Z. Hasan, Science {\bf 347}, 294 (2015).

\bibitem{XQSun} X.-Q. Sun, S.-C. Zhang, and Z. Wang, Phys. Rev. Lett. {\bf 115}, 076802 (2015).

\bibitem{HosurCRP} P. Hosur and X. Qi, C. R. Phys. {\bf 14}, 857 (2013); A. A. Burkov, J. Phys.: Condens. Matter {\bf 27}, 113201 (2015), and references therein.

\bibitem{Hughes} S. T. Ramamurthy and T. L. Hughes, Phys. Rev. B {\bf 92}, 085105 (2015).

\bibitem {CME} A. A. Zyuzin and A. A. Burkov, Phys. Rev. B {\bf 86}, 115133 (2012); J.-H. Zhou, H. Jiang, Q. Niu, and J.-R. Shi, Chin. Phys. Lett. {\bf 30}, 027101 (2013); M. M. Vazifeh and M. Franz, Phys. Rev. Lett. {\bf 111}, 027201 (2013); S. A. Parameswaran, T. Grover, D. A. Abanin, D. A. Pesin, and A. Vishwanath, Phys. Rev. X {\bf 4}, 031035 (2014);
M.-C. Chang and M.-F. Yang, Phys. Rev. B {\bf 91}, 115203 (2015); S. Zhong, J. Orenstein, and J. E. Moore, Phys. Rev. Lett. {\bf 115}, 117403 (2015).

\bibitem{NMR} D. T. Son and B. Z. Spivak, Phys. Rev. B {\bf 88}, 104412 (2013); A. A. Burkov, Phys. Rev. Lett. {\bf 113}, 247203 (2014); H.-Z. Lu, S.-B. Zhang, and S.-Q. Shen, Phys. Rev. B {\bf 92}, 045203 (2015).

\bibitem{ChiralHall}
Q. Jiang, H. Jiang, H. Liu, Q. Sun, and X. Xie, Phys. Rev. Lett. {\bf 115}, 156602 (2015);
S. A. Yang, H. Pan, and F. Zhang, \emph{ibid}. {\bf 115}, 156603 (2015).

\bibitem{BrahlekBiInSe} M. Brahlek, N. Bansal, N. Koirala, S.-Y. Xu, M. Neupane, C. Liu, M. Z. Hasan, and S. Oh, Phys. Rev. Lett. {\bf 109}, 186403 (2012).

\bibitem{ZKLiuNaBi} Z. K. Liu, B. Zhou, Y. Zhang, Z. J. Wang, H. M. Weng, D. Prabhakaran, S.-K. Mo, Z. X. Shen, Z. Fang, X. Dai, Z. Hussain, and Y. L. Chen, Science {\bf 343}, 864 (2014).

\bibitem{BorisenkoCdAs} S. Borisenko, Q. Gibson, D. Evtushinsky, V. Zabolotnyy, B. B\"{u}chner, and R. J. Cava, Phys. Rev. Lett. {\bf 113}, 027603 (2014).

\bibitem{WengTaAs} H. Weng, C. Fang, Z. Fang, B. A. Bernevig, and X. Dai, Phys. Rev. X {\bf 5}, 011029 (2015).

\bibitem{TaAsXu} S.-Y. Xu, I. Belopolski, N. Alidoust, M. Neupane, C. Zhang,
    R. Sankar, S.-M. Huang, C.-C. Lee, G. Chang, BaoKai Wang,
    G. Bian, H. Zheng, D. S. Sanchez, F. Chou, H. Lin, S. Jia, and
    M. Z. Hasan, Science {\bf 349}, 613 (2015).

\bibitem{TaAsLv} B. Q. Lv, H. M. Weng, B. B. Fu, X. P. Wang, H. Miao, J. Ma, P. Richard, X. C. Huang, L. X. Zhao, G. F. Chen, Z. Fang, X. Dai, T. Qian, and H. Ding, Phys. Rev. X {\bf 5}, 031013 (2015).

\bibitem{NbAs} S.-Y. Xu, N. Alidoust, I. Belopolski, C. Zhang, G. Bian, T.-R. Chang, H. Zheng, V. Strokov, D. S. Sanchez, G. Chang, Z. Yuan, D. Mou, Y. Wu, L. Huang, C.-C. Lee, S.-M. Huang, B. Wang, A. Bansil, H.-T. Jeng, T. Neupert \emph{et al}., Nat. Phys. {\bf 11}, 748 (2015).

\bibitem{NbP} C. Shekhar, A. K. Nayak, Y. Sun, M. Schmidt, M. Nicklas, I. Leermakers, U. Zeitler, Y. Skourski, J. Wosnitza, Z. Liu, Y. Chen, W. Schnelle, H. Borrmann, Y. Grin, C. Felser, and B. Yan, Nat. Phys {\bf 11}, 645 (2015).

\bibitem{TaPXu} N. Xu, H. M. Weng, B. Q. Lv, C. Matt, J. Park, F. Bisti, V. N. Strocov, D. Gawryluk, E. Pomjakushina, K. Conder, N. C. Plumb, M. Radovic, G. Autès, O. V. Yazyev, Z. Fang, X. Dai, G. Aeppli, T. Qian, J. Mesot, H. Ding \emph{et al}., arXiv:1507.03983.

\bibitem{YbMnBi} S. Borisenko, D. Evtushinsky, Q. Gibson, A. Yaresko, T. Kim, M. N. Ali, B. Buechner, M. Hoesch, and R. J. Cava, arXiv:1507.04847.

\bibitem{QAH} C.-Z. Chang, J. Zhang, X. Feng, J. Shen, Z. Zhang, M. Guo, K. Li, Y. Ou, P. Wei, L.-L. Wang, Z.-Q. Ji, Y. Feng, S. Ji, X. Chen, J. Jia, X. Dai, Z. Fang, S.-C. Zhang, K. He, Y. Wang \emph{et al}., Science {\bf 340}, 167 (2013); X. Kou, S.-T. Guo, Y. Fan, L. Pan, M. Lang, Y. Jiang, Q. Shao, T. Nie, K. Murata, J. Tang, Y. Wang, L. He, T.-K. Lee, W.-L. Lee, and K. L. Wang, Phys. Rev. Lett. {\bf 113}, 137201 (2014); J. G. Checkelsky, R. Yoshimi, A. Tsukazaki, K. S. Takahashi, Y. Kozuka, J. Falson, M. Kawasaki, and Y. Tokura, Nat. Phys. {\bf 10}, 731 (2014); C.-Z. Chang, W. Zhao, D. Y. Kim, H. Zhang, B. A. Assaf, D. Heiman, S.-C. Zhang, C. Liu, M. H. W. Chan, and J. S. Moodera, Nature Materials {\bf 14}, 473(2015).

\bibitem{Mahan} G. D. Mahan, \emph{Many-Particle Physics}, 3rd ed. (Springer, New York, 2007).

\bibitem{footnote} For realistic systems with multiple pairs of Weyl nodes, the full RKKY interaction contains the contributions from both the intrapair and the interpair processes.
Since the momentum spacing between different pairs is usually large, the contributions from interpair processes oscillate fast.
Thus, the contributions from the intrapair processes will dominate. The total RKKY interaction becomes a direct summation of those of different pairs.
In our Rapid Communication, we focus on this case. In fact, these contributions from the interpair processes can also be calculated in a similar mean.

\bibitem{QinLiuTI} Q. Liu, C.-X. Liu, C. Xu, X.-L. Qi, and S.-C. Zhang, Phys. Rev. Lett. {\bf 102}, 156603 (2009).

\bibitem{JJZhuTI} J.-J. Zhu, D.-X. Yao, S.-C. Zhang, and K. Chang, Phys. Rev. Lett. {\bf 106}, 097201 (2011).

\bibitem{Biswas} R. R. Biswas and A. V. Balatsky, Phys. Rev. B {\bf 81}, 233405 (2010).

\bibitem{SM} See Supplemental Material for calculation of the Green's function in the energy-coordinate representation,
derivation of the expressions of the range functions in the RKKY interaction for both the Dirac and Weyl SMs,
and relation of the RKKY interaction between the isotropic Dirac or Weyl SMs and the anisotropic ones.

\bibitem{AHEYang} K.-Y. Yang, Y.-M. Lu, and Y. Ran, Phys. Rev. B {\bf 84}, 075129 (2011).

\bibitem{Jungwirth} T. Jungwirth, J. Sinova, J. Ma{\v s}ek, J. Ku{\v c}era, and A. H. MacDonald, Rev. Mod. Phys. {\bf 78}, 809 (2006).

\bibitem{Kogan} E. Kogan, Phys. Rev. B {\bf 84}, 115119 (2011).

\bibitem{DMInt} I. Dzyaloshinsky, J. Phys. Chem. Solids {\bf 4}, 241 (1958); T. Moriya, Phys. Rev. {\bf 120}, 91 (1960).

\bibitem{Beating} The length of beating for each term is roughly determined by sine or cosine function of
$2 \zeta$ for the Fermi energy $\varepsilon_{F}=15 Q_0$ and of $2\xi_{F}$ for $\varepsilon_{F}=0.1Q_0$.
The behavior of the beating can be gained from the asymptotic expressions~\cite{SM}.

\bibitem{GXu}  G. Xu, H. Weng, Z. Wang, X. Dai, and Z. Fang, Phys. Rev. Lett. {\bf 107}, 186806 (2011).

\bibitem{ZJW1} Z. Wang, Y. Sun, X.-Q. Chen, C. Franchini, G. Xu, H. Weng, X. Dai, and Z. Fang, Phys. Rev. B {\bf 85}, 195320 (2012).

\bibitem{ZJW2} Z. Wang, H. Weng, Q. Wu, X. Dai, and Z. Fang, Phys. Rev. B {\bf 88}, 125427 (2013).

\bibitem{RKKY3D} M. A. Ruderman and C. Kittel, Phys. Rev. {\bf 96}, 99 (1954); T. Kasuya, Prog. Theor. Phys. {\bf 16}, 45 (1956); K. Yosida, Phys. Rev. {\bf 106}, 893 (1957).

\bibitem{JHZPlasmon} J. Zhou, H.-R. Chang, and D. Xiao, Phys. Rev. B {\bf 91}, 035114 (2015).

\bibitem{Vahid} Mir Vahid Hosseini and M. Askari, arXiv:1510.03020.

\bibitem{YiZhou} J.-H. Sun, D.-H. Xu, F.-C. Zhang, and Y. Zhou, Phys. Rev. B {\bf 92}, 195124 (2015). 

\bibitem{Mitchell} A. K. Mitchell and L. Fritz, Phys. Rev. B 92, 121109 (2015).

\bibitem{Abrikosov} A.A. Abrikosov, L.P. Gorkov and I.E. Dzyaloshinski, Methods of Quantum Field Theory in Statistical Physics, (Dover Publications, New York, 1963).
%
\end{thebibliography}
\end{document}